\documentclass[fleqn,twoside]{article}
\usepackage{espcrc2}

\usepackage{graphicx}
\usepackage[figuresright]{rotating}
\usepackage{amsfonts}
\newcommand{\eq}{\begin{equation}}
\newcommand{\en}{\end{equation}}
\newcommand{\ear}{\begin{eqnarray}}
\newcommand{\rae}{\end{eqnarray}}
\newcommand{\Z}{\mathbb{Z}}
\newcommand{\bra}{\langle}
\newcommand{\ket}{\rangle}

\newcommand{\AmS}{{\protect\the\textfont2
  A\kern-.1667em\lower.5ex\hbox{M}\kern-.125emS}}

\title{Confinement-deconfinement and universal string effects \\ 
from random percolation}

\author{ Ferdinando Gliozzi\thanks{Speaker at the conference}, Marco Panero and Antonio Rago \address[torino]
{Dipartimento di Fisica Teorica, Universit\`a di Torino and INFN,
sezione di Torino, via P. Giuria, 1, I-10125 Torino, Italy.}}

\begin{document}

\begin{abstract}

The 't Hooft criterion leading to confinement out of a percolating
cluster of central vortices  suggests defining  a novel
three-dimensional gauge theory directly on a random percolation process. 
Wilson loop is viewed as a counter of topological linking with
the random clusters. Beyond the percolation threshold large Wilson
loops decay with an area law and show the universal shape effects 
due to  flux tube fluctuations. Wilson loop correlators define
a non-trivial  glueball spectrum.
The crumbling of the percolating cluster when one periodic direction
narrows accounts for the finite temperature deconfinement, which
belongs to 2D percolation universality class.

\vspace{1pc}
\end{abstract}

\maketitle

\section{INTRODUCTION}
Understanding  confinement in gauge theories is a major
challenge of particle physics. Center vortices are believed to play
an important role in explaining this phenomenon. 
Long time ago 't Hooft proposed a criterion for confinement 
\cite{tHooft:1977hy}  based on three main ingredients: assuming that
 {\sl i)} there is a percolating cluster of central vortices,
 {\sl ii)} the Wilson loops measure the linking with the vortex lines 
 and  {\sl iii)} the vortices  at different places 
are weakly correlated, the sought after area decay law follows. 

The flux of a vortex is conserved modulo $N$, where
$N$ is the number of elements of the center of the gauge group. This
implies that {\sl a)} the vortices are closed lines and that 
{\sl b)} the coordination number of the intersection points is a 
multiple of $N$. While the former property  is essential for defining
topological linking,  the latter does not
take part in the argumentation.

The unsolved difficulty in order to demonstrate confinement  is
to replace the numerical evidence of the weakness of the correlation
among central vortices  with a convincing proof.

This suggests to reverse the argument using random percolation 
to define weakly correlated loops and regarding them as the central
vortices of a suitable gauge theory.
 
We define the following purely geometric setting:
generate a sample of possible states {$\{C_1,C_2,\dots\}$}
 simply by populating each of the links of a 3D lattice 
{\sl independently} with occupation probability  $p$.
The  physical observables of this system, that we call still 
{Wilson operators} $W_\gamma$, are associated to arbitrary loops 
$\gamma$ of the dual lattice with the following rule 
\begin{enumerate}
\item {$W_\gamma(C_i)=1$} if no cluster of the
  configuration $C_i$ is topologically linked to $\gamma$;
 \item {$W_\gamma(C_i)=0$} otherwise.
\end{enumerate}
The vacuum expectation value of this operator is defined by 
\eq
\bra W_\gamma\ket=\lim_{n\to\infty}
\sum_{i=1}^nW_\gamma(C_i)/n~~.
\en
Note that only the closed paths of occupied links 
 can contribute to $W_\gamma$, while the dangling ends  
do not play any role, thus in this theory the center vortices have to
be identified with the loops of the random clusters.
 
If $p$ is greater than the percolation threshold $p_c$ there is a
percolating cluster in the infinite lattice and large Wilson loops obey
the area law by construction.

The numerical implementation of this system is straightforward by
comparison with usual simulations of ordinary gauge systems: no
Markov process is needed to perform importance sampling and there are 
no thermalization problems and no critical slowing down.

On the theoretical side, both the partition function $Z$ and the gauge
group are trivial: $Z\equiv1$, $G=\Z_1$.
Some obvious questions arise:  {\sl i)} does the model have a 
well-defined continuum limit?
{\sl ii)} do the Wilson loops  exhibit the L\"uscher term and 
the other universal shape effects like in ordinary gauge theories? 
{\sl iii)} does the theory have a non-trivial glueball spectrum? 
{\sl iv)} is it possible to define a {\sl finite temperature}
deconfinement transition in the same percolation picture? 
The answer to all these questions is affirmative.

\section{ THE STRING TENSION}
\begin{figure}[htb]
\includegraphics[width=8.cm]{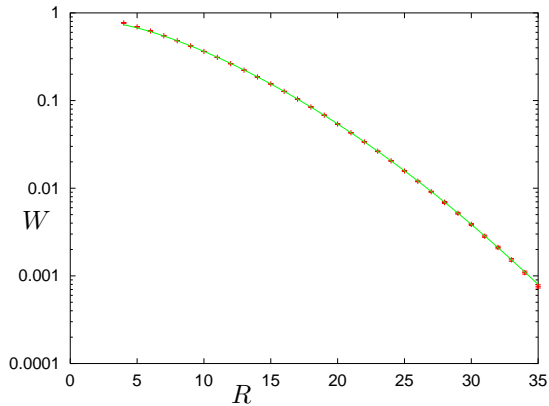}
\put(-127,4){$R$}
\put(-206,70){$W$}
\vskip -.2 cm
\caption{Expectation values of square Wilson loops $W(R)$ in a $64^3$ lattice
  at $p=0.26>p_c$ as a function of $R$. The solid line 
is a fit to Eq.(\ref{sqw}).}
\label{figure:1}
\end{figure}

We estimated the string tension $\sigma$ by fitting the mean values of
the Wilson loops associated to squares of side $R$ to the function

\eq
\bra W( R)\ket=a\,R^{\frac14}\,\exp(-b\,R-\sigma\,R^2)~~.
\label{sqw}
\en
The fits for not too small $R$ are very good (see Fig.\ref{figure:1}).

If the model under study has a well-behaved continuum limit, the
scaling form of $\sigma$ near $p_c$ should be

\eq
\sigma(p)=\sigma_o\,(p-p_c)^{2\nu}~~,
\label{sigma}
\en
where  $p_c=0.2488126(5)$ on the cubic lattice \cite{lz} and
$\nu$ is the correlation length critical exponent of  3D
percolation. We used the value $\nu=0.8765(16)(2)$
\cite{Ballesteros:1998zm}. 
In the range $0.258\le p\le 0.270$ a one-parameter fit yields  
$\sigma_o=8.90(3)$ with  $\chi^2/d.o.f\sim 0.4$. The fit is reported 
in Fig.{\ref{figure:2}.

\begin{figure}[htb]
\includegraphics[width=8. cm]{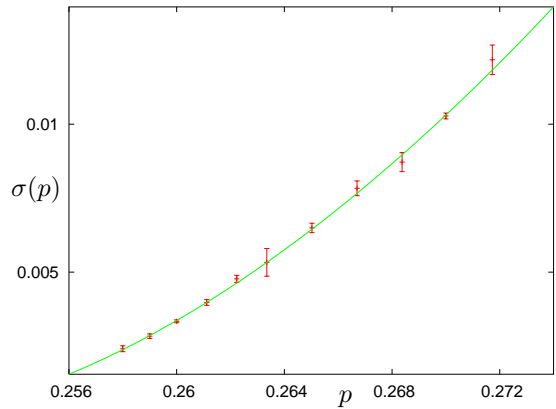}
\put(-92,2){$p$}
\put(-216,80){$\sigma(p)$}
\vskip -.3 cm
\caption{The string tension as a function of $p$. The  line
  is a one-parameter fit to Eq.(\ref{sigma}).}
\label{figure:2}
\end{figure}

Disregarding the ``string'' factor $R^{\frac14}$ in $\bra W\ket$
yields much worse fits. This suggests that the percolation process
could also account  for the  universal shape corrections ascribed to
effective string fluctuations.

A suitable quantity which is sensible to these  effects 
is \cite{Caselle:1996ii}
\eq
f=
\exp(-n^2\sigma)\frac{\bra W(L-n,L+n)\ket}{\bra W(L,L)\ket}~~,
\label{ratio}
\en
which asymptotically (large $L$ and $L-n$) should be,
in the effective string picture, a known function $f(t)$ of the ratio 
{$t=\frac nL$} without any adjustable parameters \cite{Caselle:1996ii}.
Fig.\ref{figure:3} shows a nice agreement to this conjecture.  
\begin{figure}[htb]
\includegraphics[width=8.cm]{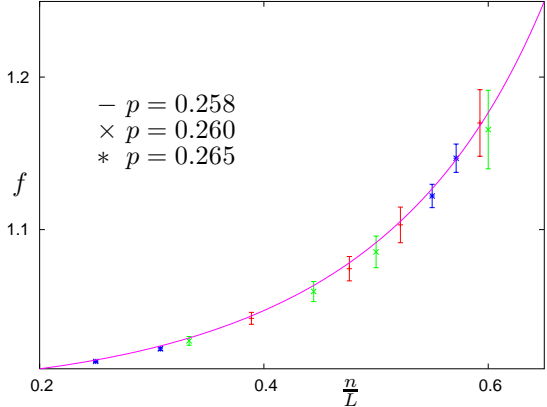}
\put(-88,2){$\frac nL$}
\put(-212,80){$f$}
\put(-180,110){$-~p=0.258$}
\put(-180,100){$\times~p=0.260$}
\put(-180,90){$*\;~p=0.265$}
\vskip -.3 cm
\caption{The quantity defined in Eq.(\ref{ratio}) for three different
  values of $p$. The line accounts for the universal shape effects
  due to effective string vibrations. No adjustable parameter is involved.} 
\label{figure:3}
\end{figure}
\section{GLUEBALLS}

Though the occupied bond connected
  correlator is exactly zero by construction, the correlator 
 among the occupied bonds \underline{belonging to a loop}} is non-zero, 
because of the constraint of being part of a closed path. The
  exponential decay rate of this correlator gives an estimate of the
  mass of the lowest scalar glueball state. We performed two numerical 
experiments at $p=0.26$ and $p=0.27$. The results reported in 
Tab.\ref{table:1} show  that the deviation from the expected scaling
  behavior is rather small.
   
\begin{table}
\vskip -.3 cm
\caption{Mass of the lowest glueball state}
\label{table:1}  

\renewcommand{\tabcolsep}{2pc} 
\renewcommand{\arraystretch}{1.2} 
\begin{tabular}{@{}lll}
\hline
$p$&$m\,a$&$m/\sqrt{\sigma}$\\
\hline
0.26&0.2189(2)&3.75(3)\\
0.27&0.3870(5)&3.81(2)\\ 
\hline
\end{tabular}
\end{table}

\section{DECONFINEMENT AT FINITE T}

 Though the usual argument for deconfinement at finite $T$ based on
 spontaneous breaking of  center symmetry here is inapplicable, 
  because there is no
 way to break $\Z_1$, it is possible to show
 that the system under study goes through a continuous, finite
 temperature, deconfining transition.

The argument runs as follows. 
Consider a lattice endowed  with a finite temperature
geometry $L\times L\times 1/T$ with fixed $p>p_c$ and start to vary $T$.
At low temperature ($L\sim 1/T$) there is a percolating cluster, so
the system is confining. As the temperature increases, the system
becomes more and more  dominated by the 2D geometry. The crucial point 
is now to observe that the percolation threshold is a decreasing
function of the space dimension.  This implies that there is a
critical value  $T(p)\gg1/L$ where the percolating
cluster crumbles away and the system  no longer confines. Clearly this
 transition belongs to the universality class of 2D percolation. 

There is now an efficient Monte Carlo algorithm for studying
percolation on any lattice \cite{nz}. It allows to calculate wrapping 
probabilities over the entire range of $p$ in a single run.
We applied this algorithm to evaluate the wrapping probability around 
the large $L$ directions.
We chose $L=40,50,64,128$ and $1/T=6$ or $1/T=8$.
Finite size scaling allows us to extrapolate the results to  $L\to
  \infty$ limit, where wrapping  probability
  is equal to percolation threshold. The results are reported
 in Tab.\ref{table:2}.
\begin{table}
\caption{Percolation threshold $p(T)$ for $1/T=6$ and 8}
\label{table:2}
\renewcommand{\tabcolsep}{2pc} 
\renewcommand{\arraystretch}{1.2} 
\begin{tabular}{@{}lll}
\hline
$\frac1{T}$&$p(T)$&$T_c/\sqrt{\sigma}$\\
\hline
6&0.272355(5)&1.497(4)\\
8&0.265615(5)&1.510(4)\\ 
\hline
\end{tabular}
\end{table}

\end{document}